\documentclass[twocolumn,aps,showpacs,showkeys,amsmath,amssymb,floatfix]{revtex4}
\usepackage{amsfonts}
\usepackage{graphicx}
\usepackage{amsfonts}
\usepackage{graphicx}

\newcommand{\wb}{\Omega_{\mathrm{b}}}
\newcommand{\no}{\nu_{\mathrm{o}}}
\newcommand{\wo}{\omega_{\mathrm{o}}}
\newcommand{\nb}{\nu_{\mathrm{b}}}
\newcommand{\ks}{k_{\mathrm{s}}}
\renewcommand{\k}{\kappa}
\newcommand{\m}{m^*}
\newcommand{\KT}{\mathrm{K_BT}}
\newcommand{\totalfig}{0.48\textwidth}

\begin{document}

\title{Moving breathers in bent DNA with realistic parameters}

\author{J Cuevas}
\address{University of Sevilla. Group of Nonlinear Physics,
Departamento de F\'{\i}sica Aplicada I. c/. Virgen de Africa 7,
41011-Sevilla (Spain)}
\author{EB Starikov}
\address{Karolinska Institute, Center for Structural
Biochemistry NOVUM, S\,-\,14157 Huddinge, Sweden}
\author{JFR Archilla}
\address{University of Sevilla. Group of Nonlinear Physics,
Departamento de F\'{\i}sica Aplicada I. Avda. Reina Mercedes, s/n.
41012-Sevilla (Spain)}
\author{D Hennig}
\address{Freie Universit\"{a}t Berlin. Fachbereich Physik,
Institut f\"{u}r Theoretische Physik\\ Arnimallee 14, 14195
Berlin, Germany}

\begin{abstract}
Recent papers have considered moving breathers (MBs) in DNA models
including long range interaction due to the dipole moments of the
hydrogen bonds. We have recalculated the value of the charge transfer
when hydrogen bonds stretch using quantum chemical methods which takes
into account the whole nucleoside pairs. We explore the consequences
of this value on the properties of MBs, including the range of frequencies
for which they exist and their effective masses. They are able to
travel through bending points with fairly large curvatures provided
that their kinetic energy is larger than a minimum energy which depends
on the curvature. These energies and the corresponding velocities
are also calculated in function of the curvature.
\end{abstract}

\maketitle

\section{Introduction}
\label{sec:intro}

Breathers are localized oscillations in coupled networks of nonlinear
oscillators \cite{MA94}. They have been extensively studied in the
last years \cite{A97,FW98,PHYSD99,M00,CHAOS03}. Under some conditions
they can move easily along the network while maintaining their localization
\cite{CAT96,AC98}. A physical system where they may play a significant
role is DNA, because the hydrogen bonds between nucleotides are highly
nonlinear and their openings can be related to biological processes
such as transcription, replication and, in the case of MBs, with the
transport of information, energy and charge \cite{PF00,P04}.

DNA is also a flexible chain, and the question arises of to what extent
MBs are hindered in their movement by the points of bending, or if,
on the other hand, these points may have a biological role by trapping
breathers and accumulating their energy.

The shape of the DNA molecule is felt because the hydrogen bond is
a polar one, and the dipolar interaction between the dipoles is a
long-range one which depends on the distances and orientations of
the dipoles, and, therefore, on the shape.

Soliton properties in the Discrete Nonlinear Schr\"odinger
Equation  framework with dipole-dipole interaction have been
considered in Refs. \cite{Chris98,GMCR97}, among others, whereas
MBs in DNA Klein-Gordon models have been studied in
Ref.~\cite{CAGR02} and their interaction with bending points in
Ref.~\cite{CPAR02}. A problem for these studies is the fairly
large number of physical parameters that are not well known, among
them, crucially, the coefficient of dipole interaction. In this
paper we present new quantum chemical calculation that lead to the
value of this parameter and investigate its consequences.
Moreover, the reduction of the number of parameters allow us to
explore the variation of another one, the breather frequency,
which previously was considered as fixed.

\section{Quantum chemical calculation of the charge transfer}

\label{sec:quantum}

The only quantity that it is needed to implement the dipole--dipole
interaction is the \emph{charge transfer} $q$. It is defined in the
following way: if $p_{0}$ is the dipole moment of a Watson-Crick
hydrogen bond at the equilibrium distance, and it is stretched by
a small amount $u$, the new dipole moment is given to the first order
in $u$ by $p=p_{0}+q\, u$. The relevant estimations have been obtained
in Ref.~\cite{CAGR02}, yielding $q$ values between $-0.0014\,\mathrm{e}$
and $-0.0183\,\mathrm{e}$ for an A-T base pair and between $-0.025\,\mathrm{e}$
and $-0.055\,\mathrm{e}$for a G-C base pair, depending on the quantum-chemical
method used. However, these values were obtained only for Watson-Crick
base pairs \emph{in vacuo}. In the present work we significantly extend
our model, in that we take into account

a) the variation of the dipole moment of the whole nucleoside (base
+ deoxyribose) pair,

b) the influence of the DNA duplex environment on dipole moments
of the nucleoside pairs in question.

We also consider the helical structure of the DNA double strand.
 The procedure now is as
follows: we consider regular homogeneous stacked trimers of
nucleoside pairs, namely, adeninosine-thymidine (AT) and
guanosine-cytidine (GC) pairs in the following arrangement:
AT/AT/AT and GC/GC/GC. We take the 'effective' dipole moment of a
nucleoside base pair to be one third of the trimer dipole moment.
Then we stretch and squeeze H-bonds by gradually adding or
subtracting of up to 0.1 Angstrom to the equilibrium spacing
between A and T (or G and C) in the central pair of the trimer,
with the two flanking pairs remaining in the standard
B-DNA-conformation. At each of the H-bond stretching/squeezing
states, the 'effective' dipole moment of a nucleoside pair was
estimated using semiempirical quantum chemistry. The relevant
computational details are published elsewhere \cite{HSAP04}.

We evaluate a linear regression of the 'effective' dipole moment ($p$)
onto the nucleotide spacing change ($u$) in the Watson-Crick pairs
($p=p_{0}+q\, u$), where the coefficient $q$, delivers the desired
estimate. As a result, we have obtained $p_{0}=1.84\,\mathrm{D}$,
$q=-0.09\,\mathrm{e}$, for the AT pair and $p_{0}=3.15\,\mathrm{D}$,
$q=-0.10\,\mathrm{e}$, for the GC pair. The Watson-Crick H-bond stretching
diminishes the effective dipole moment of nucleoside pairs, which
ought to be connected with the proper changes in the purine-pyrimidine
molecular orbital overlap included into quantum-chemical evaluations
of dipole moments. If we take into account only the conventional electrostatic
charge distributions, increasing the spacing between the partners
in the Watson-Crick base pair should lead to the increase in the dipole
moment. Note that the values of $q$ obtained in this way are substantially
larger than the ones obtained in Ref.~\cite{CAGR02} and very similar
for both the A-T and G-C pairs, in spite of the different number of
hydrogen bonds.
%newstarikov
 This could be explained by the appreciable influence
of the DNA duplex surrounding on the Watson-Crick A-T and G-C base
pair dipole moments. We agree that taking an "effective" dipole
moment of the stacked base pair trimer is only an approximation.
We intended to include the coupling between the nearest neighbours
in the DNA stack. That the "effective" dipole moment values differ
from those of the isolated base pairs is actually a significant
result showing that we were right to include the DNA stacking
interactions at least approximately into our model. As to the
Coulomb electron-electron interaction, it was considered at our
Hartree-Fock-level quantum-chemical calculations in the form of
Pauli correlations, where electrons of the same spin are
repellent. Note also that being the values of $q$ for both
homopolynucleotide duplexes very similar we should not expect $q$
to change significatively for heterogeneous DNA.

\section{The model}

\label{model}

The Hamiltonian system also used in Ref.~\cite{CAGR02,CPAR02} is
a Peyrard--Bishop model~\cite{PB89} augmented with long-range interaction.
It can be written as:

\begin{eqnarray}
H=\sum_{n=1}^{N}\Big(\frac{1}{2}m\dot{u}_{n}^{2}+D\,(e^{-b\, u_{n}}-1)^{2}+\nonumber \\
\frac{1}{2}k\,(u_{n+1}-u_{n})^{2}+\frac{1}{2}\sum_{i}J_{in}u_{i}\,
u_{n}\Big). \label{ham}
\end{eqnarray}

The explanation of the terms and variables is as follows: $u_{n}$
is the stretching of the $n$-th hydrogen bond; $m$ is the reduced
mass of a nucleotide pair; the Morse potential $D\,(e^{-b\,
u_{n}}-1)^{2}$, represents the energy of a hydrogen bond between a
pair of nucleotides; $\frac{1}{2}k\,(u_{n+1}-u_{n})^{2}$ is the
stacking energy between neighboring bonds;
$\frac{1}{2}\sum_{i}J_{in}u_{i}\, u_{n}$ is the energy of the
dipole--dipole interaction; if $\vec{r}_{n}$ is a vector that
denotes the position of each particle in the 2-d space,
$J_{in}=J/|\vec{r}_{n}-\vec{r}_{i}|^{3}$, for $i\neq n$ and $0$
otherwise, where $J$, hereafter referred to as the dipole
parameter, is given by $J=q^{2}/(4\,\pi\epsilon_{0}\, d^{3})$, $q$
and $d=3.4$~\AA, being the charge transfer and the distance
between neighbouring base pairs, respectively, as deduced in
Ref.~\cite{CAGR02}. We approximate the shape of the molecule in
the vicinity of a bending point by a parabola with curvature
$\kappa$ embedded in a plane, and consider the dipoles with the
same orientation and orthogonal to it. We consider fixed geometry
as the H--bond vibrations are much faster than the bending
movement. Thus, $\vec{r}_{n}=(x_{n},y_{n})$ with $y_{n}=\kappa
x_{n}^{2}/2$. We also neglect the heterogeneity of DNA.

%%%%%%%%%%%%%%%%%figure1%%%%%%%%%%%%%%%%%%%%%%
\begin{figure}
\begin{center}
\includegraphics[width=\totalfig]{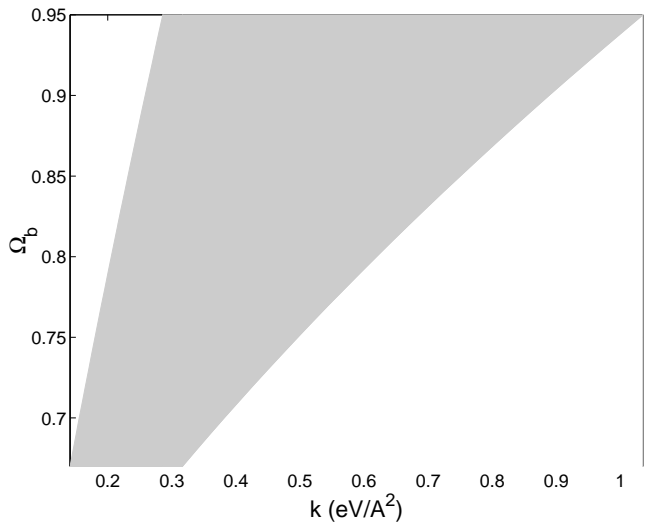}
\end{center}
\caption{Range of existence of moving breathers. The upper limit
corresponds to the bifurcation of stability inversion. The lower
limit is the top frequency of the phonon band. The dispersion is
smaller if closer to the upper limit.} \label{fig:stabbif}
\end{figure}

The values of the parameters are taken from Ref.~\cite{CAGR02},
i.e., $D=0.04$~eV, $b=4.45$~\AA$^{-1}$ and $m=300$~amu. As
discussed in the same reference, $k$ can take values between $0.01$~eV/\AA$^{2}$
and $10$~eV/\AA$^{2}$ and, in consequence, we consider it as an
adjustable parameter in that range. The value of $J$ for a charge
transfer $q$ intermediate between the A-T and G-C base pairs is $J=0.0031$~eV/\AA$^{2}$.

\section{Moving breathers in a straight chain}

Breathers, exact to machine precision, are calculated numerically
using techniques based in the anti-continuous limit~\cite{MA96}.
MBs with good movability properties, i.e., low dispersion, are
obtained for values of the parameters in the vicinity of a
bifurcation point, called \emph{inversion of stability}, where a
single breather becomes unstable and a double one stable, or vice
versa, and perturbing them with a vector collinear to the
eigenvector that becomes unstable~\cite{CAT96,AC98}.
%%%%%%%%%%%%%%%%%%%%%%%%%figure2%%%%%%%%%%%%%%%
\begin{figure}
\begin{center} \includegraphics[width=\totalfig]{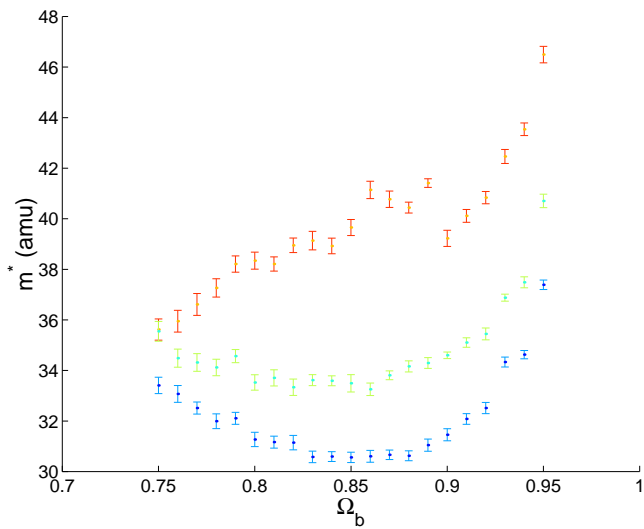}\end{center}
\caption{Dependence of the effective mass $\m$ with respect to the
breather frequency for three different values of the coupling
constant $k$: 0.2376, 0.2693 and 0.3010 eV/\AA$^{2}$ from bottom
to top.} \label{fig:mass}
\end{figure}
%%%%%%%%%%%%%%%%%%%%%%%%%figure3%%%%%%%%%%%%%%%%%%%%%%
\begin{figure}
\begin{center}
\includegraphics[width=\totalfig]{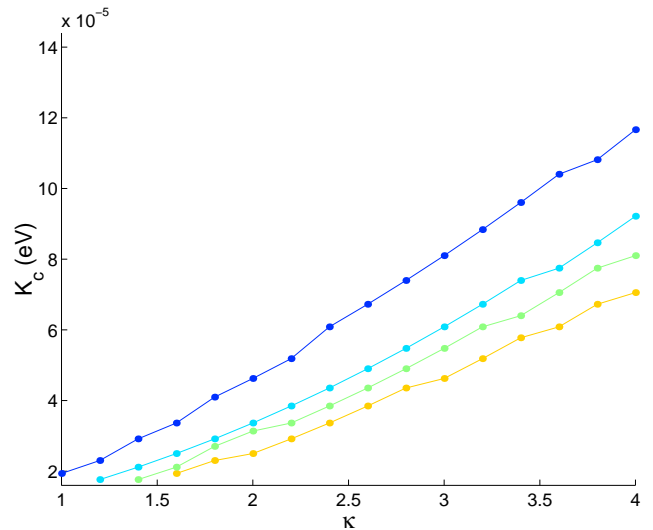}
\end{center}
\caption{Dependence of the critical translational energy $K_{c}$
with respect to $\k{}$ for $k=\ks$. The frequencies $\wb$ of the
breathers are (from up to down) 0.95, 0.90, 0.85, and 0.80.
Smaller values of $\k{}$ are not shown as $K_{c}$ is very small
and no accurate results can be obtained.} \label{fig:kappaen}
\end{figure}

%%%%%%%%%%%%%%%%%%%%%%%%%%%%%%figure4
\begin{figure}
\begin{center}
\includegraphics[width=\totalfig]{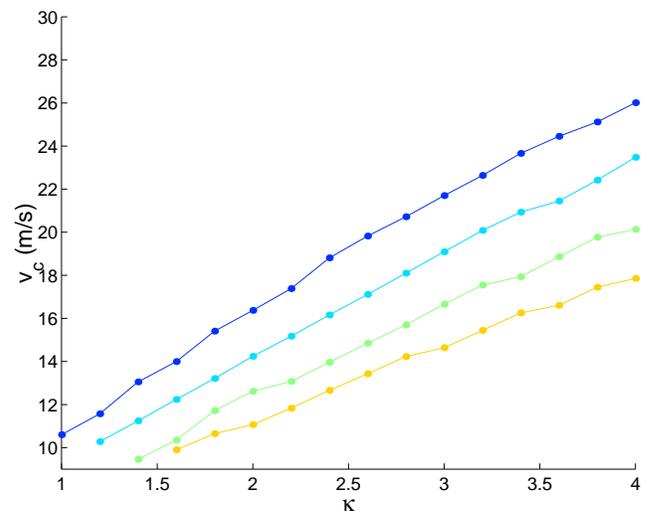}
\end{center}
\caption{Same as Fig.~\ref{fig:kappaen} but for the critical
initial translational velocity ($v_{c}$).} \label{fig:kappavel}
\end{figure}

The particles of the breather oscillate coherently with a frequency
$\nb$. For the sake of simplicity, we define the non-dimensional
frequency $\wb=\nb/\no$, where $\no$, defined through \begin{equation}
\no=\frac{1}{2\pi}\sqrt{\frac{2b^{2}D}{m}}=1.13\,\,\textrm{THz}\,\end{equation}
 is the linear frequency of the isolated oscillators in Eq.~\ref{ham}.
As the on--site potential is soft, $\wb<1$. We have limited its
values to the interval $(0.65,0.95)$ because lower frequencies
brings about breathers difficult to move, due to their small
width, and MB with larger ones develop frequencies which interact
with the phonon band. The latter is due to the fact that MBs are a
wave packet with frequencies around the one of the perturbed
static breather.

For a given value of the breather frequency $\wb$, MBs exist for
values of $k$ above a critical one $\ks=\ks(\nb)$, for which the
inversion of stability takes place. This value has to be calculated
numerically.

The phonon band is composed of the frequencies of the linear modes,
given by~\cite{CAGR02,GMCR97}:%
\begin{eqnarray}
\Omega_{ph}^{2}=\left(\frac{\omega_{ph}}{\wo}\right)^{2}=
1+\frac{4k}{m\,\wo^{2}}\,\sin^{2}\frac{q}{2}+\frac{2J}{m\,\wo^{2}}F_{3}(\Re(e^{iq})),
\end{eqnarray}%
 where $q$ is the wave number, $\wo=2\,\pi\,\no$ and $F_{s}(z)=\sum_{k>1}z^{k}/k^{2}$
is the Polylogarithmic or Jonqui\`{e}re function. They exist for
increasing values of $k>\ks$ until an upper limit $k_{m}=k_{m}(\nb)$,
for which the top of the phonon band, at $q=\pi$, reaches the second
harmonic of the breather. $k_{m}$ is given by \begin{equation}
k_{m}=\left(\wb^{2}-\frac{1}{4}\right)\, m\,\wo^{2}+\frac{3}{8}\, J\,\zeta(3)\,,\end{equation}
 where $\zeta(s)$ is the Riemann's zeta-function. This is also the
upper limit for MBs existence. Fig.~\ref{fig:stabbif} shows the
range of existence of MBs. It can be seen that they exist for a wide
range of frequencies, but they have much smaller dispersion at the
vicinity of the inversion stability curve and we will restrict often
our study to them.

Static breathers in the above mentioned region have an energy
between $\KT$ and $3.7\ \KT$, with $T=310\ $K. The dipolar energy of
these breathers oscillates between $0.006\ \KT$ and $0.023\ \KT$.
The energy depends monotonically with the frequency, corresponding
the highest energy to the smallest frequency.

Moving breathers behave as quasiparticles, having an effective mass
$\m$ related with the kinetic energy of the breather by
$K=\frac{1}{2}\m\, v^{2}$, with $v$ being the translational velocity
of the breather~\cite{CAT96,AC98}. $K\in(1.5,50)\times10^{-4}\ \KT$
is also the kinetic energy added to the static breather. $\m$ is
approximately constant as long as $v$ is small enough and,
therefore, emits small phonon radiation. Fig.~\ref{fig:mass} shows
the dependence of the mass with respect to the frequency for three
different values of $k$.

Thus, MBs have an effective mass $\m$ between 30 and 46~amu,
depending on the value of frecuency and the stacking parameter.

\section{Moving breathers in bent chains}

In order to study the effect of the curvature on breather mobility,
we launch a breather through the bending point. The bending acts as
a potential barrier~\cite{CPAR02} in accordance to the trapping
hypothesis formulated in Ref.~\cite{CPAR02b} (for a discussion of
this last point see Ref.~\cite{AGCC02}). It implies that breathers
are reflected when reaching the bending point as long as their
translational energy $K$ is below a critical point $K_{c}$.
Fig.~\ref{fig:kappaen} shows the dependence of this value with
respect to $\k{}$ for several frequencies. The dependence of the
critical initial velocity is also shown in Fig.~\ref{fig:kappavel}.
$K_{c}$, which is also the amount of energy given to a static
breather to start its movement, is fairly small compared to the
thermal energy at physiological temperature of ~0.027~eV. Therefore,
it seems likely that it can be given easily by the environment, and
most MBs are able to pass through bending points without being
reflected.

\section{Conclusion}

In this paper we consider a model for DNA with long--range
interaction due to the dipole moments of the hydrogen bonds among
nucleotides. The dipole parameter $J$ has been obtained through
quantum chemical calculations including the whole nucleoside. The
consequences for moving breathers in straight and bent DNA chains
have been analyzed. These can be summarize as: 1) For each
different value of the stacking parameter there exist MBs with
different frequencies, including an upper limit, determined
numerically for the inversion of stability curve, and a lower
limit determined both numerically and analytically by the top of
the phonon band. 2) MBs breather have a fairly large effective
mass of 30--50~amu. 2) The values of $J$, though small, are large
enough for the MBs to feel the the bending of DNA. 3) We have
calculated the minimum kinetic energy and the velocity of a
breather to be able to pass over a bending point in function of
the curvature. As the minimum kinetic energy is small, it seems
plausible that most MBs will have enough energy to pass through
bending points, although the ones with low energy will be
reflected.

\section*{acknowledgments}

This work has been partially support under the European Commission
RTN project LOCNET, HPRN-CT-1999-00163 and the MECD--FEDER project
BMF2003-03015/FISI. JC acknowledges an FPDI grant from `La Junta de
Andaluc\'{\i}a'.

\newcommand{\noopsort}[1]{} \newcommand{\printfirst}[2]{#1}
  \newcommand{\singleletter}[1]{#1} \newcommand{\switchargs}[2]{#2#1}

\end{document}